# Challenges of Securing Massively Multiplayer Online Games

Kolten Sinclair[1], Steven Womack[2], Jacob Elliott[3], Benjamin Stafford[4], Sundar Krishnan[5]

Angelo State University, San Angelo, Texas, United States

[1] ksinclair@angelo.edu , [2] jelliott22@angelo.edu , [3] swomack5@angelo.edu , [4] bstafford2@angelo.edu , [5] skrishnan@angelo.edu

*Abstract*—When it comes to security in the modern world, things have improved a lot since the early 2000s. Hypertext Transfer Protocol Secure (HTTPS) and Transport Layer Security (TLS) have made the transfer of our data across the internet much safer than years prior, and the advent of VPNs and private browsing have only compounded that. However, the gaming industry has been notoriously behind the curve when it comes to security, most notably with Massively Multiplayer Online (MMO) games, which due to the intrinsic nature of their architecture, have an astounding amount of ground to cover. In this paper, the authors discuss the challenges that MMO developers face when trying to design a secure game, as well as some more modern approaches to security that will help improve the industry moving forward. The authors also highlight a few real-life examples of exploits and breaches that have happened and look at how they were mitigated.

## I. INTRODUCTION

Securing any online system is no simple task, but combine that with the games industry's chaotic development culture, and the job becomes extremely difficult to get right. This means that security practices and challenges in this field are always evolving, and there is a constant stream of new methods and procedures for securing game data in every aspect of the application. The topic in question, however, is how we continue to improve upon systems that seem to work well in practice, especially in the context of a large online system like an MMO online game.

MMO games, also referred to as Massively Multiplayer Online games, are among the most popular video game genres. Multiplayer online games establish a whole ecosystem within the game, from having a richly drawn world to having a large player population that engages in interactions and completes missions with one another to have a working economy that benefits every player equally. An MMO online game allows hundreds or thousands of players to communicate at the same time in a gaming world that they are connected to over the Internet [1]. This type of game is usually played in a persistent online multiplayer setting. MMO online games provide a persistent universe in which the game persists even when no one else is playing. Few of these games feature any meaningful single-player elements or client-side artificial intelligence because they either heavily or entirely focus on multiplayer gaming [1]. Players cannot, therefore, "beat" MMOs in the manner that single-player games are often played. The free-to-download MMO "Star Sonata" can be in which the player assumes the command of a spacecraft and can become the "Emperor" and "win" the game by dealing and negotiating their way up to the top. Nevertheless, most MMOs have many non-player (or non-playable) characters (NPC) and Mobs that either provide tasks or act as opponents.

## II. LITERATURE REVIEW

Due to MMOs and game security being a niche genre up until the last decade or so, there are only a few research papers and projects to pull data from. Regardless, the few that we've found have been extensive and informative, especially highlighting some of the more technical details associated with their research. Currently, there is a great focus on the network portion of MMO security, primarily because most exploits are developed in that layer of the game. Gavrić et al. [2] state that Distributed Denial of Service (DDoS) attacks are some of the most prevalent forms of malicious activity, typically used to cause server outages, with some cases losing the company's serious revenue. Summeren [3] also states that the other main form of network exploitation comes in the form of cheat engines and packet alteration. Overall, the landscape for MMO security research is sparse, but it looks like that will be changing in the next few years with the introduction of new technologies.

## III. DISCUSSION

There are a lot of moving parts in a fully-fledged MMO, which can make security a daunting task. This doesn't just start while the game is running, however. It also takes place during the game's development.

### A. Development

Release timelines for new content make it exceedingly difficult to accurately decide what will and will not be in the final release patch, which means that there are constantly new features being added, other features being changed, and some even being deprecated altogether. With all these changes happening at a rapid pace due to quarterly expectations, code can often be left in the source that is not meant to be there. Rarely do these snippets of code contain vulnerable functions and/or behaviors, and upon discovery by malicious players can be catastrophic if not caught quickly. This issue is only compounded by the constantly evolving codebase of an MMO, in which developers often release frequent content patches and updates, leaving plenty of opportunity for the scenario to present itself. There are many different security recommendations to follow when creating software, which itself has its own field. However, one quite common mistake developers can make is writing code that is vulnerable to overflow attacks. These attacks work by sending more data than a function expects, therefore breaking its intended behavior and causing unexpected side effects. These effects display the need for a dedicated and funded quality assurance team that can work in sync with the development team to find these exploits.

## B. Complexity

Along with the security of the code itself, there is also another consideration to be made, which is how these different segments of the codebase communicate with each other. The customer base for MMOs requires that new content be added on a regular basis, and with each new module added to the source, that makes the overall behavior of the game more complex, giving more opportunity for failure within these 'inter-game' communications.

Another facet that adds to the complexity of MMO is the large number of events that make up these 'inter-game' communications. Fischer et al. [4] break down MMO events into multiple dimensions, such as context, persistency, synchronization, validity, delivery, and security. Since guaranteeing cheat-free operations is prohibitively expensive, a company must determine the level of security per individual event that a user may perform.

Complexity isn't isolated just to the game's inner workings, but it's also a problem when the client and the server communicate and synchronize information. Games like MMOs demand low-latency transfer of data so that the game client can be updated with new information from the server in near-real time. This requirement makes it much harder to ensure that these connections are completely secure and prevents many types of standard verification methods. Due to this, MMO companies will often leave themselves vulnerable to various kinds of network-based exploits.

## C. Network

The communication between the client and the server is the most important piece of an MMO. Thus, it's also the most targeted area for exploitation by malicious players. Typically, the servers hosting these games will accept any valid connection request, regardless of whether it is meant to be malicious or not. This opens the hosting servers to possible DDoS/DOS attacks, as well as more direct access attacks from people trying to access information stored on the game servers. Before the traffic gets to the server, however, it must come from the client, and this connection between them was commonly targeted in the early days of the gaming industry. One older method of attack is packet sniffing/alteration. This essentially allows the attacker to observe the inbound and outbound packet traffic with a commercial tool such as Wireshark or tcpdump, intercept this traffic, and make changes in transit so that when it gets to the game server, it has been altered in a way to give the player an advantage or another desired outcome. Thankfully, due to modern encryption standards such as TLS, attacks like these are much less common according to [5]. Even though the communication of the data itself has gotten more secure, there are still ways people have found to break the game's communications differently. McGraw et al. [6] explain how specifically timed events and actions can have a very detrimental effect on the game server. For example, if a player drops an item on the ground and immediately logs out, it's possible for the server to see the item on the ground. However, because the player left so fast the server did not update their inventory, so when the player logs back, the item will be on the ground and still in their inventory. This has been a common method for duplicating items in any online game for quite a while but is made much easier by the sheer amount of information that needs to be moved back and forth between a client and server in an MMO. In another vein, Hilven et al. [7] go more in-depth into how a specific MMO, World of Warcraft, successfully implemented the BitTorrent protocol for use in their client update procedures, which acts to securely download update files from other players who may be closer to the end user than the official game servers, therefore decreasing server load and connection issues.

## D. Client

The client is the most targeted portion of any MMO because it is the piece that the end-user has the most control over, making it the most vulnerable to a plethora of different attacks. One of the most observed exploits is the automation of in-game actions through the use of external programs or scripts. As discussed by Efe et al. [5] and McGraw et al. [6], such programs can have a wide array of implementations, some of which modify the game client itself, which poses a huge security risk for the game server. Dynamic Link Libraries (DLLs) are arbitrary modules of code that can be inserted into the game client's memory stack to modify in-game values before they are sent to the server for processing. In modern games, however, this method is becoming much more difficult to use as developers get better at locking down their game clients and verifying information discrepancies between what the client has sent and what the server currently has on record. To get around this, exploit developers have now moved to more stealthy approaches, typically in the form of third-party applications that don't attempt to modify the game client at all but rather use the information given on the player's screen to carry out game tasks automatically by-passing simulated inputs to the client.

## E. Fraud

One major consideration to be made when designing a game is monetization. Many titles will allow players to buy in-game currency and items, and any time real money is brought into the equation, there are sure to be malicious players who will try to take advantage of it. Hilven et al. [7] discuss how it's exceedingly difficult to correlate real-world and in-game transactions, so a customary practice in MMOs is to create an account of high value and then sell it through third-party marketplaces such as eBay or Craigslist. Ordinarily, this would be against TOS (terms of service) regardless of whether the account's value was obtained legitimately or not, but with the addition of using external tools to help achieve such high-valued accounts much easier, it becomes a real problem for any kind of in-game economy to try and remain unpolluted with 'dirty' money.

Another form of dirty money that may appear in the virtual economy is currency through the organization of gold farmers and money launderers. A gold farmer is a user who focuses exclusively on conducting in-game actions that produce resources such as items or currency. They often use TOS-breaking software to maximize efficiency. Hogben et al. [8] Money launderers are those who trade/or gift the in-game currency to users and receive real-world money back. Since the real-world transaction is not recorded, the security team will instead need to keep a watch for users with a history of suspicious trades. Some existing methods that current-day companies use are human interactive proofs like CAPTCHA human observational proofs that come from reports, as well as data mining techniques [10].

There is also the consideration that some games allow players to purchase items directly in-game with real-world currency. If such games have any kind of undocumented duplication glitch, it can be extremely easy for that feature to destroy the in-game economy as well as quarterly profits by flooding the in-game market with high-value items all at once. Maintaining an MMO's economic system is a constant battle against bot accounts, duplication glitches, and other exploits, so gamemasters (the people in charge of administrating the game servers while in-game) must be exceedingly vigilant when it comes to sniffing out bad actors.

*F. Players and Social Engineering*

A game's security can only prevent so much, and in the case of MMOs, they unfortunately provide the ideal environment for social engineers and malicious players to thrive. Because MMOs rely on a large player base to operate at a profit, this means that there is a constant amount of harassment and bullying that goes on accordingly. Some of the specific types of harassment include ganking (repeatedly killing another player), kill stealing (getting the final hit on an enemy and stealing the loot/experience from it), ninja looting (taking more loot that a player was meant to during a group fight), and griefing (targeting other players and causing damage to their character and/or possessions). This harassment doesn't have to be physical either, however, as things such as hurtful language in voice/chat communications can be just as detrimental to a player's experience.

Due to the competitive nature of many MMOs, another form of harassment that can also manifest is malicious and false reporting of previously listed behaviors. Malicious reporting can be especially dangerous if the security team overly relies on automated systems for their reporting. Improper banning and suspension of innocent users will lead to brand damage as well as users permanently leaving the game. The possibility of a large user group leaving increases if leaders of user-led community groups are suspended during critical in-game events such as weekly resource gathering events or new content launch times.

Another thing to be wary of when it comes to the safety of players is the possibility of social engineering. While most MMO's TOS states that players should be above a certain age to play, that doesn't always happen, and even with that, most of them allow younger children to play too. These players are frequently targeted by other players attempting to steal their login information and more sinister behaviors like grooming and stalking.

*G. Mitigating Exploitation*

As MMOs evolve, so will the exploits and their developers, so the security personnel need to stay on top of any new advancements in the field as best they can. Over the years, multiple techniques have come about to aid in the fight against bad actors, such as memory checksums, timing audits, and quick patching. However, with new threats ever-presently looming on the horizon, the need for newer methods of security grows. One possible topic for discussion is to find new ways to distribute the hosting of the game servers. In a typical MMO hosting structure, a centralized network of servers is used to host the world, maintain connections, etc.

Kabus et al. [11] offer several alternatives to the traditional client/server architecture and instead look at how costs and scaling may be offloaded to the users by moving to a P2P (Peer-to-peer) structure. These approaches are Mutual Checking, Log Auditing, and Trusted Computing. Mutual Checking utilizes the concept of relying on the consensus of multiple unaffiliated clients. Goodman et al. [12] go into further detail about a potential model that can be used for Mutual Checking, which utilizes a message resolution as well as a trust system to decide upon disciplinary action. Log Auditing would utilize specific users to hold a record of events until they log off, after which the record will be audited, and any cheating will be rolled back as much as possible. Trusted computing would be a way to find and verify users that can be trusted to host server information.

Another possible approach would be to use a distributed system to allow all this heavy processing to be distributed amongst a large array of devices that are able to easily scale their computations and move them to other systems before they become overloaded. This is very important, as overloaded servers often cause network/game outages and other game-breaking bugs, which are all exploitable by bad actors. Improbable is a new company specializing in distributed computing for games, and it's very likely that they could hold the key to the next generation of MMO gaming and security.

IV. CONCLUSION

Overall, security is becoming more and more of a concern for MMO developers, and the safety and privacy of its players are more important than ever. Unfortunately, the amount of research in the field is lacking, but with the release of newer, more complex games, that should begin changing very soon, and the industry will evolve with it. Moving forward, an emphasis on more secure server operations would be the logical next step, considering the large amount of DDoS attacks and server outages that happen in modern MMOs.

REFERENCES

[1] Massively multiplayer online game, https://wowpedia.fandom.com/wiki/Massively_multiplayer_online_game

[2] N. Gavrić and Ž. Bojović, "Security Concerns in MMO Games—Analysis of a Potent Application Layer DDoS Threat," Sensors, vol. 22, no. 20, p. 7791, Oct. 2022, doi: 10.3390/s22207791.

[3] R. Van Summeren, "Security in online gaming," RADBOUD UNIVERSITY NIJMEGEN, 2011. Accessed: Nov. 27, 2022. [Online]. Available: https://www.cs.ru.nl/bachelors-theses/2011/Rens_van_Summeren___0413372___Security_in_Online_Gaming.pdf

[4] T. Fischer, M. Daum, F. Irmert, C. Neumann, and R. Lenz, "Exploitation of event-semantics for distributed publish/subscribe systems in massively multiuser virtual environments," Proceedings of the Fourteenth International Database Engineering & Applications Symposium on - IDEAS '10, 2010, doi: 10.1145/1866480.1866494.

[5] A. Efe and E. Önal, "ONLINE Game Security: A Case Study of an MMO Strategy Game," Gazi University Journal of Science, vol. 7, no. 2, pp. 43–57, Jul. 2020, Accessed: Nov. 27, 2022. [Online]. Available: https://dergipark.org.tr/en/download/article-file/649301

[6] G. McGraw and G. Hoglund, "Online Games and Security," IEEE Security & Privacy Magazine, vol. 5, no. 5, pp. 76–79, Sep. 2007, doi: 10.1109/msp.2007.116.

[7] A. Hilven and A. Woodward, "How safe is Azeroth, or, are MMORPGs a security risk?," Australian Information Security Management Conference, Dec. 2007, doi: 10.4225/75/57b548f5b875a.

[8] G. Hogben et al., "Security and privacy in massively-multiplayer online games and social and corporate virtual worlds," research.utwente.nl, Nov. 2008, Accessed: Nov. 28, 2022. [Online]. Available: https://research.utwente.nl/en/publications/security-and-privacy-in-massively-multiplayer-online-games-and-so

[9] S. Chun, D. Choi, J. Han, H. K. Kim, and T. Kwon, "Unveiling a Socio-Economic System in a Virtual World," Proceedings of the 2018 World Wide Web Conference on World Wide Web - WWW '18, 2018, doi: 10.1145/3178876.3186173.

[10] K. Woo, H. Kwon, H. Kim, C. Kim, and H. K. Kim, "What can free money tell us on the virtual black market?," Proceedings of the ACM SIGCOMM 2011 conference on SIGCOMM - SIGCOMM '11, 2011, doi: 10.1145/2018436.2018484.

[11] P. Kabus, W. W. Terpstra, M. Cilia, and A. P. Buchmann, "Addressing cheating in distributed MMOGs," Proceedings of 4th ACM SIGCOMM workshop on Network and system support for games - NetGames '05, 2005, doi: 10.1145/1103599.1103607.

[12] J. Goodman and C. Verbrugge, "A peer auditing scheme for cheat elimination in MMOGs," Proceedings of the 7th ACM SIGCOMM Workshop on Network and System Support for Games - NetGames '08, 2008, doi: 10.1145/1517494.1517496.